# Deterministic quantum correlation in an interferometric scheme

Byoung S. Ham
Center for Photon Information Processing, School of Electrical Engineering and Computer Science, Gwangju Institute of Science and Technology
123 Chumdangwagi-ro, Buk-gu, Gwangju 61005, South Korea
bham@gist.ac.kr
(submitted on Dec. 17, 2020)

**Abstract:**
Over the last several decades, entangled photon pairs generated from $\chi^{(2)}$ nonlinear optical materials via spontaneous parametric down conversion processes have been intensively studied for various quantum correlations such as Bell inequality violation and anticorrelation. In a Mach-Zehnder interferometer, the photonic de Broglie wavelength has also been studied for quantum sensing with an enhanced phase resolution overcoming the standard quantum limit. Here, the fundamental principles of quantumness are investigated in an interferometric scheme for controllable quantum correlation not only for bipartite entangled photon pairs in a microscopic regime, but also for macroscopic coherence entanglement generation.

**Introduction**

Quantum entanglement [1] is the heart of quantum technologies such as quantum computing [2-7], quantum communications [8-10], and quantum sensing [11-14]. Although intensive research has been performed for quantum correlation in interferometric schemes of a Hong-Ou-Mandel (HOM) dip [15-19], photonic de Broglie wavelength (PBW) [20-25]), Bell inequality violation [26-28], and Franson-type nonlocal correlation [29-33]), the fundamental physics of entangled photon-pair generation itself has still been vailed in terms of the coincidence measurement probability. Thus, nondeterministic measurement-based quantum technologies are still extremely inefficient compared with their classical counterparts. The classical technologies are of course deterministic as well as macroscopic.

Recently, a novel method of deterministic quantum correlation has been proposed [34-36] and experimentally demonstrated [37,38] to unveil secretes of quantum entanglement for both HOM dip and PBW using pure coherence optics. Thus, on-demand entangled photon pair generation can be manipulated and applied for deterministic quantum information. Based on coherence optics, collective control of quantum features is a great benefit for future quantum technologies compatible with the classical counterparts. Here, the fundamental physics of coherence-based quantum correlation is investigated to unveil secretes of quantumness in an interferometric scheme of both anticorrelation and nonlocal correlation. For $\chi^{(2)}$ −generated entangled photon pairs [39], some misunderstandings of quantum correlation are also pointed out to specify the secret of nonclassical nature.

Figure 1 shows a particular scheme of quantum correlation in an interferometric scheme, where entangled photon pairs are generated from a $\chi^{(2)}$ nonlinear material via spontaneous parametric down conversion (SPDC) processes [39]. Due to the spontaneous emission, however, entangled photon pairs should have a random phase relation in the coincidence detection measurements, resulting in coherence modification [40]. Thus, the success probability of higher-order entangled photon pair generation severely decreases as the order (i.e., entangled photon number) increases [21-25]. In this context, quantum applications of PBW for quantum sensing [25] and photonic quantum computing [2] have become impractical. Most significantly, the fundamental question of how to generate entangled photon pairs has not been answered clearly. Unlike most approaches in quantum correlation studies based on the particle nature of photons, we use the wave nature of photons to investigate the basic physics of quantumness in an interferometric scheme of a coupled Mach-Zehnder interferometer (MZI).

Assuming there is a suitable phase relation between the signal $(s)$ and idler $(i)$ photons in each entangled pair generated from a $\chi^{(2)}$ nonlinear medium via SPDC processes, the basic equations for coincidence detection measurements in Fig. 1 can be derived by using general matrix representations of pure coherence optics: $\begin{bmatrix} E_\alpha \\ E_\beta \end{bmatrix} = [BS1][\zeta]\begin{bmatrix} E_s \\ E_i \end{bmatrix}$; $[BS1] = \frac{1}{\sqrt{2}}\begin{bmatrix} 1 & i \\ i & 1 \end{bmatrix}$; $[\zeta] = \begin{bmatrix} e^{i\zeta} & 0 \\ 0 & 1 \end{bmatrix}$ [34]. The input field $E_s$ ($E_i$) represents the signal (idler) photons, where the number of entangled photons per pair is nondeterministically decided by the nonlinear optical process. Coherent treatment for the entangled photons originates from the intrinsic properties assigned by Heisenberg's uncertainty principle in terms of the energy-time relation in the $\chi^{(2)}$ process: $\Delta f \Delta t \geq 1$. Here, a 50/50 balanced nonpolarizing beam splitter (BS) plays an important role not only to test but also to create the quantum features in output ports via superposition of two input photons [41].

In Fig. 1, the first (second) MZI is controlled by $\Delta L_1$ ($\Delta L_2$), where $\zeta = \frac{2\pi}{\lambda}\Delta L_1$, $\varphi = \frac{2\pi}{\lambda}\Delta L_2$, and $\lambda$ is the degenerate input wavelength of the photon pair composed of s and i. The output ports $\alpha$ and $\beta$ of BS1 are coincidently measured for photon correlation, where the normalized coincidence detection rate $R_{\alpha\beta}$ is proportional to $\langle I_\alpha I_\beta \rangle$, and $I_\alpha = I_0(1 + \sin\zeta)$ and $I_\beta = I_0(1 - \sin\zeta)$ are the intensities corresponding to each output port of $\alpha$ and $\beta$, respectively. Thus, the coincidence detection rate is given by:

$$R_{\alpha\beta} = \frac{1}{2}(1 + \cos 2\zeta), \tag{1}$$

resulting in the anticorrerlation condition $(R_{\alpha\beta} = 0)$ as $\zeta = \pm(m - \frac{1}{2})\pi$ (m = 1,2,3,...). This is the hidden secrete of the entangled photon pair of s and i, where there must be a $\frac{\pi}{2}$ (or $\frac{\lambda}{4}$) phase difference between the paired photons [34]. This relation is equivalent to that between the sine and cosine functions in a Hilbert space of a bipartite system. In other words, the origin of the entanglement of bipartite entities corresponds to the orthonormal bases of the system [36]. Another example satisfying equation (1) is given by perpendicularly polarized photon pairs under coherence optics [24].

In most HOM dip observations, however, the expected $\zeta$ ($\lambda$) −based sinusoidal oscillation by equation (1) has never been observed. Very recently, such a $\lambda$ −dependent coherence feature in coincidence detection measurements has been proposed [40] and experimentally demonstrated using coherent photons [38]. As analyzed in ref. 40, the disappearance of $\lambda$ −dependent $g^{(1)}$ coherence in most HOM dip measurements is due to coherence washout (incoherence) by spontaneous emission-caused random phases given to entangled photon pairs, resulting in a modified $\zeta$ in the mean value of coincidence measurements [40]. Due to this averaging effect, the anticorrelation condition of $\zeta$ $(= \pm\frac{\pi}{2})$ is replaced by the overall coherence-washout relation, $\Delta_G \tau \geq 1$, where $\Delta_G$ is the spectral bandwidth of the $\chi^{(2)}$ −generated entangled photons [15,39]. In other words, the disappearance of $g^{(1)}$ coherence in conventional HOM dips is not due to the no-wave nature of single photons but instead due to the intrinsic properties of $\chi^{(2)}$ −generated entangled photon pairs with random phases. Note that the $\zeta$ −like coherence feature appearing in the intensity correlation of a polarization-based system is similar to the present case in terms of orthogonal bases under the Fresnel-Arago law [42]: $e^{i\varphi} \rightarrow e^{i\xi_{sp}}$, where $\xi_{sp}$ is an angle between the perpendicularly polarized paired photons.

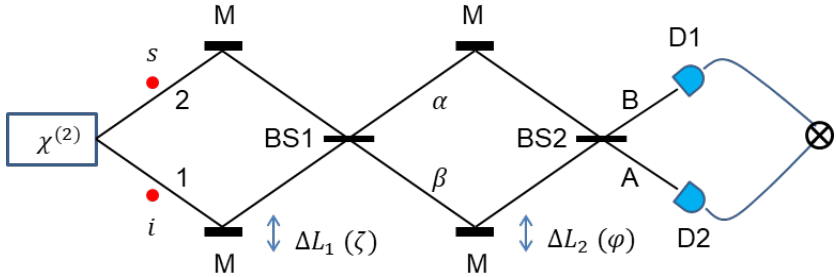

Fig. 1. Schematic of entanglement generation [21]. BS: beam splitter, D: detector, M: mirror. $\zeta = \frac{2\pi}{\lambda}\Delta L_1$; $\varphi = \frac{2\pi}{\lambda}\Delta L_2$.

Coherent photon pairs s and i satisfying equation (1) can be easily generated from any commercially available laser using a BS [37]. Then, the $\chi^{(2)}$ −generated entangled photon pairs interfering on a BS are exactly equivalent to the coherent light in an MZI if anticorrelation in equation (1) is satisfied [34]. Thus, the anticorrelation profile now include the $g^{(1)}$ coherence term, resulting in controllable quantum features with $\zeta$.

This discovery of coherently controllable quantum features unveils the quantumness of the wave nature. In other words, coincidence measurements are not the bottom line for quantum features, but instead an extreme case of coherence optics satisfying the uncertainty relation [40]. Thus, $\zeta$-dependent quantum correlation opens the door to deterministic and macroscopic quantum science, because a BS or MZI does not differentiate whether the input is a single photon or not [34,43].

For the coincidence detection measurements in Fig. 1, the following amplitude relations are obtained:

$$\begin{bmatrix} E_A \\ E_B \end{bmatrix} = \frac{1}{2} E_0 \begin{bmatrix} e^{i\varphi} - 1 & ie^{i\zeta}(1 + e^{i\varphi}) \\ i(1 + e^{i\varphi}) & e^{i\zeta}(1 - e^{i\varphi}) \end{bmatrix}. \quad (2)$$

From equation (2), the corresponding intensities are as follows:

$$I_A = I_0(1 + \cos\zeta \sin\varphi), \quad (3)$$
$$I_B = I_0(1 - \cos\zeta \sin\varphi), \quad (4)$$

where $I_0 = E_0 E_0^*$. Thus, the normalized coincidence detection rate, $R_{AB} = I_A I_B / I_0^2$, becomes (see Fig. 2(a)):

$$R_{AB} = 1 - (\cos\zeta \sin\varphi)^2. \quad (5)$$

For the anticorrelation condition with $\zeta = \pm\frac{\pi}{2}$ for the first MZI as given in equation (1), the output photons in equations (3) and (4) show no fringe or quantum features at all, regardless of $\varphi$ ($= \frac{2\pi}{\lambda}\Delta L_2$), as shown in Figs. 2(a) and 2(b) (see the dashed lines). This is because the second BS (BS2) equally splits the bunched photons achieved by BS1. On the contrary, a maximal photon bunching effect ($R_{AB} = 0$) as a quantum feature is accomplished by $\zeta = \pm m\pi$, where m=0,1,2... Unlike the discussions in ref. 21, quantum features cannot be obtained in Fig. 1 without a shift of $\zeta$. Moreover, a PBW with $\lambda_B = \lambda/2n$ cannot be achieved either, unless higher-order entangled photons are involved, whereas $\lambda_B$ is known to be a direct result of interference between two or more entangled photon pairs [22-25]. Although PBWs have been well studied, details on how to generate PBWs from individual (independent) photons are not clearly known. As numerically calculated in Figs. 2(c) and 2(d), the quantum determinacy is controllable in an interferometric scheme with coherent photons [34-38,40].

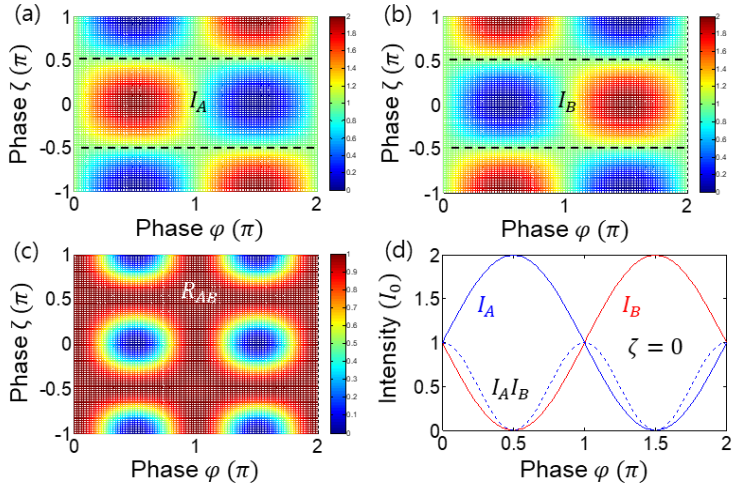

Fig. 2. Numerical simulations for Fig. 1. (a) and (b) Intensities $I_A$ in path A and $I_B$ in path B. (c) coincidence detection rate $R_{AB}$. (d) Output intensity for $\zeta = 0$. Dotted: $R_{AB}$.

Recently, a coherent $\lambda_B$ has been proposed [35] and experimentally demonstrated in a coupled MZI system [37]. Unlike photon superposition-based $\lambda_B$ in conventional PBWs, a coherence version of PBW can also be created via path superposition between two or more MZIs in a modified interferometric scheme. For this, asymmetric coupling between MZIs plays an important role [36]. Figure 3 shows the modified interferometric scheme for coherence de Broglie wavelength (CBW) via path superposition between MZIs. As analyzed in Figs. 1 and 2, a BS is an intrinsic quantum device, and thus an MZI is as well [34,43].

The matrix representation of the first block for n=1 in Fig. 3 is as follows:

$$\begin{bmatrix} E_\alpha \\ E_\beta \end{bmatrix} = [\Psi][MZI] \begin{bmatrix} E_0 \\ 0 \end{bmatrix}$$

$$= \frac{1}{2}\begin{bmatrix} e^{i\psi}(1-e^{i\varphi}) & ie^{i\psi}(1+e^{i\varphi}) \\ i(1+e^{i\varphi}) & -(1-e^{i\varphi}) \end{bmatrix}\begin{bmatrix} E_0 \\ 0 \end{bmatrix}, \qquad (6)$$

where $[MZI] = \frac{1}{2}\begin{bmatrix} 1-e^{i\varphi} & i(1+e^{i\varphi}) \\ i(1+e^{i\varphi}) & -(1-e^{i\varphi}) \end{bmatrix}$, $[\Psi] = \begin{bmatrix} e^{i\psi} & 0 \\ 0 & 1 \end{bmatrix}$, and n is the number of the basic building blocks.

For n=2, the following is satisfied by the asymmetric coupling of $\psi$:

$$\begin{bmatrix} E_A \\ E_B \end{bmatrix} = \begin{bmatrix} E_\alpha \\ E_\beta \end{bmatrix}' \begin{bmatrix} E_\alpha \\ E_\beta \end{bmatrix}\begin{bmatrix} E_0 \\ 0 \end{bmatrix}$$

$$= \frac{1}{4}\begin{bmatrix} e^{i\psi}\left[(1-e^{i\varphi})^2 - (1+e^{i\varphi})^2\right] & i\left[e^{i\psi}(1+e^{i\varphi})(1-e^{i\varphi}) - (1+e^{i\varphi})(1-e^{i\varphi})\right] \\ i\left[e^{i2\psi}(1+e^{i\varphi})(1-e^{i\varphi}) - e^{i\psi}(1+e^{i\varphi})(1-e^{i\varphi})\right] & -e^{i2\psi}(1+e^{i\varphi})^2 + e^{i\psi}(1-e^{i\varphi})^2 \end{bmatrix}\begin{bmatrix} E_0 \\ 0 \end{bmatrix}, \quad (7)$$

where $\begin{bmatrix} E_\alpha \\ E_\beta \end{bmatrix}'$ is for the opposite location of $\psi$ for n=2, corresponding to $-\psi$ in $\begin{bmatrix} E_\alpha \\ E_\beta \end{bmatrix}$. As demonstrated in refs. 35 and 37, the output fields result in either an identity relation, $\begin{bmatrix} E_A \\ E_B \end{bmatrix} = (-1)\begin{bmatrix} 1 & 0 \\ 0 & 1 \end{bmatrix}\begin{bmatrix} E_0 \\ 0 \end{bmatrix}$, for $\psi = 0$ [44] or CBW for $\psi = \pm\pi$ [35]:

$$\begin{bmatrix} E_A \\ E_B \end{bmatrix} = \frac{-1}{2}\begin{bmatrix} (1+e^{i2\varphi}) & i(1-e^{i2\varphi}) \\ -i(1-e^{i2\varphi}) & (1+e^{i2\varphi}) \end{bmatrix}\begin{bmatrix} E_0 \\ 0 \end{bmatrix}. \qquad (8)$$

Thus, the nonclassical output intensities are coherently achieved as follows (see Figs. 4(a) and (b)):

$$I_A = \frac{1}{2}I_0(1+\cos 2\varphi), \qquad (9)$$

$$I_B = \frac{1}{2}I_0(1-\cos 2\varphi). \qquad (10)$$

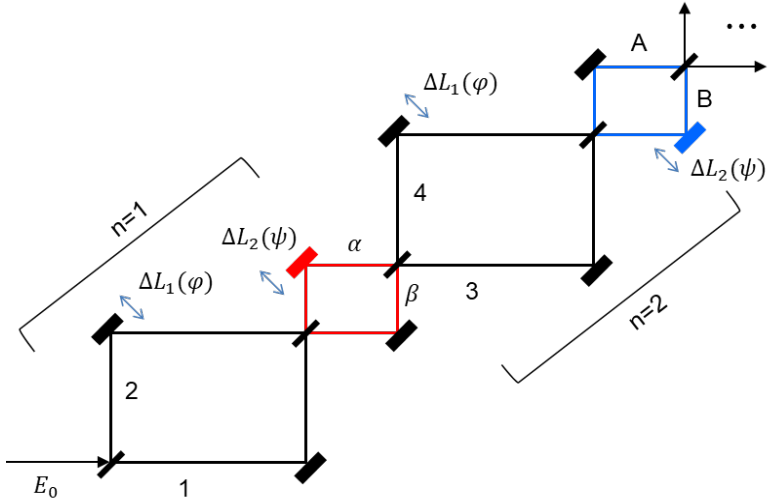

Fig. 3. Schematic of path superposition between MZIs. $\varphi = \frac{2\pi}{\lambda}\Delta L_1$; $\psi = \frac{2\pi}{\lambda}\Delta L_2$.

For an n-coupled MZI structure with an alternating $\psi$−coupling in Fig. 3 (see the colored boxes), a general solution for the CBW is obtained:

$$\begin{bmatrix} E_\alpha \\ E_\beta \end{bmatrix}^{(n)} = \frac{1}{2}(-1)^{n-1}\begin{bmatrix} (1+(-1)^n e^{in\varphi}) & i(1-(-1)^n e^{in\varphi}) \\ -i(1-(-1)^n e^{in\varphi}) & (1+(-1)^n e^{in\varphi}) \end{bmatrix}\begin{bmatrix} E_0 \\ 0 \end{bmatrix}, \qquad (11)$$

where the corresponding intensities are as follows (see Figs. 4(c) and (d) as well as the Supplementary Information):

$$I_\alpha^{(n)} = \frac{1}{2}I_0[1+(-1)^n\cos(n\varphi)], \qquad (12)$$

$$I_\beta^{(n)} = \frac{1}{2}I_0[1-(-1)^n\cos(n\varphi)]. \qquad (13)$$

In equations (12) and (13), the condition needed for anticorrelation or photon (field) bunching is $\varphi = \frac{m\pi}{n}$ (m=0,1,2,…,n), as shown in Figs. 4(e) and (f). As a result, the fundamental phase bases of the n-coupled MZI are 0 and $\pi/n$. This relation is fundamental to the macroscopic quantum feature, which is equivalent to the PBW [20-25]. As is well understood, the enhanced phase resolution results from the tensor product between n-bipartite MZIs, where PBW is used for entangled photon pairs [22], while CBW is for MZIs [36]. In both cases, superposition via the tensor product plays a critical role. The normalized coincidence detection rate $R_{AB}$ is given by $\langle I_\alpha^{(n)} I_\beta^{(n)}\rangle/\langle I_0\rangle^2$ (see Fig. 4(e) and (f) as well as the Supplementary Information):

$$R_{AB}^{(n)} = \frac{1}{2}\langle 1 - cos2n\varphi\rangle. \quad (14)$$

Thus, the phase resolution in n-coupled asymmetric MZIs is enhanced by a factor of n in an n-MZI coupled system, where n=1 is the classical limit of the Rayleigh criterion or diffraction limit (see the blue curve in Fig. 4(f)).

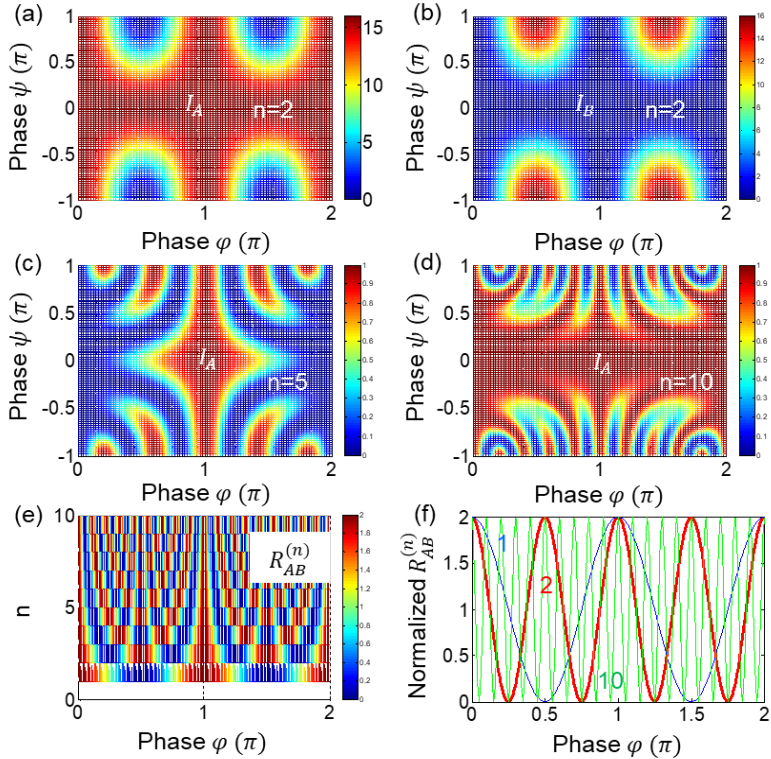

Fig. 4. Numerical simulations for $\psi = \pi$ Fig. 3. (a)-(d) Output intensity. (e) and (f) Coincidence detection rate $R_{AB}^{(n)}$. In (f), the number indicates n=1 (blue), n=2 (red), and n=3 (green) $\psi = \pi$. The unit of $I_A$ and $I_A$ is $I_0$ $(= E_0 E_0^*)$.

Considering that the intensity correlation of coherent light is $g^{(2)}(0) = 1$ as determined by Poisson photon statistics, the normalized coincidence detection rate in equation (14) is the same as the intensity correlation $g^{(2)}$. Unlike the general understanding of entangled photon pair-based nonclassical features, the quantum feature in Figs. 3 and 4 is due to MZI superposition, where the asymmetric $\psi$ plays a key role. The numerical simulations for equations (11)-(14) are shown in Fig. 4. Figures 4(a) and (b) for n=2 are compared with Figs. 2(a) and (b) for n=1, respectively, in terms of the Heisenberg limit and standard quantum limit, respectively. The enhanced phase resolution for $n \geq 2$ is direct evidence of the quantum feature based on coherence optics. Unlike PBW, CBW is deterministic. Moreover, a macroscopic feature of CBW is an inherent property of the wave nature of photons. In other words, the classical feature of the input photons (or fields) is converted into a quantum feature of CBW, as seen in Fig. 4 via alternating path superposition of $\psi$s.

In conclusion, the quantum features of anticorrelation and CBW were analyzed in an interferometric scheme of coupled MZIs using coherence optics, where asymmetric coupling method plays an essential role. Due to the coherence optics of the BS and MZI, the generated quantum feature was deterministically and collectively controlled for on-demand macroscopic quantum features of Schrodinger's cat. For the $\chi^{(2)}-$generated entangled photon pairs, a $\pi/2$ phase relation was derived from a simple matrix representation. Thus, the quantumness is now controllable collectively and coherently. The present research opens the door to coherence quantum technologies.

Acknowledgment

BSH acknowledges that this work was supported by GIST via GRI 2020.